\documentclass[aps,groupedaddress,showpacs,twocolumn]{revtex4}

\usepackage{amsmath}
\usepackage{epsfig}
\begin{document}

\preprint{HEP/123-qed}

\title[Short Title]{Completely Positive Maps with Memory}

\author{Sonja Daffer,$^1$
        Krzysztof W$\acute{\mbox{o}}$dkiewicz,$^{1,2}$
        James D. Cresser,$^3$}
\author{John K. McIver$^1$}%

\affiliation{%
    $^1$Department of Physics and Astronomy,
    University of New Mexico,
    800 Yale Blvd. NE,
    Albuquerque, NM 87131   USA  \\
    $^2$Instytut Fizyki Teoretycznej,
    Uniwersytet Warszawski, Ho$\dot{z}$a 69,
    Warszawa 00-681, Poland  \\
    $^3$ Department of Physics,
    Macquarie University,
    NSW 2109, Australia
    }%

\date{\today}      
\begin{abstract}

The prevailing description for dissipative quantum dynamics is
given by the Lindblad form of a Markovian master equation, used
under the assumption that memory effects are negligible. However,
in certain physical situations, the master equation is essentially
of a non-Markovian nature. This paper examines master equations
that possess a memory kernel, leading to a replacement of white
noise by colored noise.
The conditions under which this leads to a completely positive,
trace-preserving map are discussed for an exponential memory
kernel.
\end{abstract}


\maketitle

The theory of open quantum systems deals with systems that
interact with an environment.  This physically realistic situation
is of crucial importance in physics in general.  The overall
system-environment state is described by a density operator that
evolves unitarily. Most often, one is interested in the system
alone, which is described by a reduced density operator obtained
by tracing over the environment degrees of freedom. The evolution
of the system state is described by a completely positive map
\cite{kraus1983}.

The earliest prominent example of an open quantum system is the
Weisskopf-Wigner theory of spontaneous emission describing the
irreversible decay of an atom to its ground state.  The standard
approach is to make approximations that simplify the mathematics
without changing the underlying physics. The vast majority of the
theory is built on the Markov approximation. It is assumed that
the correlation time between the system and environment is
infinitely short so that memory effects can be neglected.  This
leads to the ubiquitous quantum Markovian master equation that
describes the time evolution of the system density operator. This
differential equation has a canonical form in terms of a Lindblad
superoperator.  The Lindblad superoperator generates a completely
positive dynamical semigroup
\cite{lindblad1976,davies1969,alicki1987}.  The solution to the
differential equation for the system density operator defines a
completely positive linear map.

Certainly, physical situations arise where correlations between
the system and environment exist for a small yet finite period of
time.  Or the Lindblad superoperator may necessarily be of a more
general form that depends on time. This complicates the theory,
which can no longer be presented in the framework of quantum
dynamical semigroups \cite{haake1973}.  To date, there lacks a
solid theory of system-environment interactions in which memory
effects are incorporated.  In this Letter, we present a model that
describes system-environment interactions with memory. Because
this occurs in all branches of physics \cite{examples0000} it is
of considerable importance to develop a general formalism that
extends beyond the Markov approximation and resulting Lindblad
evolution.

We first review the canonical form of the generator that leads to
the quantum Markovian master equation for the system density
operator. The Gorini-Kossakowski-Sudarshan generator
\cite{gorini1976} has the form
\begin{equation}     \label{eq:gks}
     L \rho=-i [H,\rho ] + \frac{1}{2} \sum^{n^2-1}_{i,j=1}
     c_{ij} \{ [F_i,\rho F_j ] + [ F_i \rho, F_j ]  \}
\end{equation}
for $\rho \in \mathcal{M}_n$ ($\mathcal{M}_n$ denotes the set of
$n \times n$ complex matrices), where $H=H^\dagger,
\textrm{Tr}(H)=0, \textrm{Tr}(F_i)=0, \textrm{Tr}(F_i
F_j)=\delta_{ij},$ and $(c_{ij})$ is a complex positive
semidefinite matrix.  The first term describes the unitary
evolution while the second term defines the Lindblad superoperator
$\mathcal{L}$ describing the dissipative dynamics due to the
interaction of the system with the environment.
The solution $\rho(t)=e^{L t}\rho(0)$ to Eq. ( \ref{eq:gks})
defines a linear operator $\Phi_t: \rho \rightarrow \rho_t$ that
maps the system density operator at some initial time to the
system density operator at some time in the future.  The
expression in Eq. (\ref{eq:gks}) has been shown to generate a
completely positive dynamical semigroup that has the following
properties:
\begin{eqnarray}   \label{eq:properties}
    &(i)& \parallel \Phi_t \rho \parallel_1 =
    \parallel \rho \parallel_1  \forall \hspace{.05in} \rho \in V_1^{+}(\mathcal{H}),
    \hspace{.05in} t\geq0,   \nonumber \\
    &(ii)& \Phi_t \otimes I_n \geq 0
    \hspace{.05in} \forall \hspace{.05in} n \in \textrm{Z}_+,  \\
    &(iii)& \textrm{lim}_{t \downarrow 0} \hspace{.05in} \Phi_t =
    I,     \nonumber\\
    &(iv)& \Phi_t \Phi_s = \Phi_{t+s}, \hspace{.05in} t,s\geq 0,
    \nonumber
\end{eqnarray}
where $\parallel \cdot \parallel_1$ denotes the trace norm in the
space of linear operators on the Hilbert space $\mathcal{H}$ and
$V_1^{+}(\mathcal{H})$ is the cone of all positive semidefinite
elements in the space \cite{kossakowski1972}.  The first property
states that the map is trace-preserving for all density operators
for all times. The second property states that the map is not only
positive, but completely positive, i.e., any extension to a larger
space remains a positive map \cite{choi1972}.  The third property
expresses the continuity at the origin and hence for all time.
Finally, \ref{eq:properties}$(iv)$ is the semigroup property. This
property arises when the environment or reservoir is
delta-function correlated in time as is the case for white noise
diffusive processes.

The prevailing description for dissipative dynamics is given by
$\dot{\rho}= \mathcal{L} \rho$, where $\mathcal{L}$ is given by
the second term in Eq. (\ref{eq:gks}). Instead, we consider
dissipative dynamics described by
\begin{equation}  \label{eq:kernelequation}
    \dot{\rho}= K \mathcal{L} \rho,
\end{equation}
where $K$ is an integral operator that depends on time of the form
$K \phi = \int_0^t k(t-t') \phi(t') \textrm{d}t'$. The kernel
function $k(t-t')$ is a well-behaved, continuous function that
determines the type of memory in the physical problem. The
solution to the master equation can be found by taking the Laplace
transform
\begin{equation}  \label{eq:laplace}
     s \tilde{\rho}(s) - \rho(0) = \tilde{K}(s) \mathcal{L}
    \tilde{\rho}(s),
\end{equation}
determining the poles, and inverting the equation in the standard
way. The solution to Eq. (\ref{eq:kernelequation}) defines a
linear map $\Phi_t: \rho \rightarrow \rho_t$ that describes the
evolution of a system coupled to an environment provided that
$\Phi_t$ satisfies properties $2(i), (ii)$, and $(iii)$.  Because
the master equation is no longer of the Lindblad form, the
semigroup property is lost. However, this property is not
necessary to describe a physically acceptable state evolution. All
that is required is that the linear map be a completely positive,
trace-preserving map.

The evolution of quantum systems is described by unitary
operators. Therefore, $\Phi_t$ should describe an evolution of the
system that arises from an overall unitary evolution of the system
and environment
\begin{equation}   \label{eq:unitary}
    \Phi_t(\rho)=\textrm{Tr}_\Gamma \{ U ( \rho \otimes
    |\gamma_0\rangle \langle \gamma_0|)U^\dagger \},
\end{equation}
where $\Gamma$ denotes the environment degrees of freedom and
$\gamma_0$ is some initial state of the environment. One should
imagine that the state $\rho$ is prepared at some time $t=0$ and
so is initially uncorrelated with the external system. The state
and the environment evolve unitarily for some time and they become
correlated.  One may think of the environment as extracting
information from the system as it will typically map pure states
into mixed states.
This noise process is described by a linear map involving only
operators on the system of interest so that it has a Kraus
decomposition
\begin{equation}
  \Phi_t(\rho) = \sum_k A_{k}^{\dagger} \rho A_k,
  \label{eq:kraus}
\end{equation}
where the condition $\sum_k A_k A^\dagger_k=I$ ensures that unit
trace is preserved for all time~\cite{kraus1983}. A map has a
Kraus decomposition if and only if it is completely positive.
Physically, the existence of a Kraus decomposition ensures that
the system evolution is compatible with a unitary evolution on the
system-environment Hilbert space.

In solving Eq. (\ref{eq:kernelequation}) it is advantageous, in
practice, to find a damping basis~\cite{briegel993} for the
superoperator $\mathcal{L}$ that diagonalizes the master equation.
Solving the eigenvalue equation ${\cal L} \rho = \lambda \rho$
produces a complete, orthogonal basis with which to expand the
density operator at any time.  This results in a set of
eigenvalues $\{ \lambda_i \}$ and right and left eigenoperators,
$\{ R_i \}, \{ L_i \},$ that satisfy the duality relation
$\textrm{Tr} \lbrace L_i R_j \rbrace = \delta_{ij}$. Once the
initial state is known $\rho(0)= \sum_{i} \textrm{Tr} \lbrace L_i
\rho(0) \rbrace R_i$, the state of the system at any later time
can be found through $\rho(t)=\sum_i \textrm{Tr} \lbrace L_i
\rho(0) \rbrace \Lambda_i(t) R_i=\sum_i \textrm{Tr} \lbrace R_i
\rho(0) \rbrace \Lambda_i(t) L_i$. The $\Lambda_i(t)$ are general
functions that determine the time evolution. The damping basis
allows for the replacement of $\mathcal{L}$ by the eigenvalues
$\lambda_i$ so that Eq. (\ref{eq:laplace}) may be written as
\begin{equation}
    \tilde{\rho}(s)=\sum_i \frac{1}{s-\tilde{K}(s) \lambda_i}
    \textrm{Tr}\{ L_i \rho(0) \}  R_i.
\end{equation}

We now present a concrete example of a system that has a master
equation of the form of Eq. (\ref{eq:kernelequation}).  Consider
the time-dependent Hamiltonian
\begin{eqnarray}
    H(t) &=& \hbar \sum_{i=1}^3  \Gamma_i(t) \sigma_i,
\end{eqnarray}
where $\Gamma_i(t)$ are independent random variables and
$\sigma_i$ are the Pauli operators.  Each random variable obeys
the statistics of a random telegraph signal, which is defined by
$\Gamma_i(t)=a_i (-1)^{n_i(t)}.$  The random variable $n_i(t)$ has
a Poisson distribution with a mean equal to $t/ 2\tau_i$, while
$a_i$ is an independent coin-flip random variable. By abuse of
notation, the random variable $a_i$ takes the values $\pm a_i$.
The random telegraph signal is a wide sense stationary stochastic
process \cite{vankampen1981} with zero mean. This model is
applicable to any two-level quantum system that interacts with an
environment possessing random telegraph signal noise. For example,
this could describe a two-level atom subjected to a fluctuating
laser field that has jump-type random phase noise
\cite{wodkiewicz1984}. In the language of nuclear magnetic
resonance, this model describes a spin-1/2 particle in the
presence of three orthogonal magnetic fields. Each field has a
constant magnitude $a_i$ and inverts randomly in time with a
distribution given by $n_i$.  The strength of the coupling of the
system to the external influence is given by the parameters $a_i$.
The flipping or fluctuation rate is inversely given by $\tau_i$.

The equation of motion for the density operator is given by the
von Neumann equation $\dot{\rho}=-\frac{i}{\hbar} [H,\rho]=-i
\sum_k \Gamma_k(t) [\sigma_k,\rho]$ that has the solution
\begin{equation}   \label{eq:integralsolution}
    \rho(t) = \rho(0)-i \int_0^t  \sum_k \Gamma_k(s) [\sigma_k,\rho(s)]
    \textrm{d}s.
\end{equation}
Upon substitution of Eq. (\ref{eq:integralsolution}) back into the
von Neumann equation and performing a stochastic average, one
obtains
\begin{equation}   \label{eq:rtsrhodot}
    \dot{\rho}(t) = - \int_0^t  \sum_k  e^{-\frac{t-t'}{\tau_k}} a_k^2
    [\sigma_k,[\sigma_k, \rho(t') ]] \textrm{d}t',
\end{equation}
where the correlation functions of the random telegraph signal
$\langle\ \Gamma_j(t) \Gamma_k(t') \rangle = a_k^2
e^{-\frac{|t-t'|}{\tau_k}} \delta_{jk}$ have been employed.
Equation (\ref{eq:rtsrhodot}) is an incoherent combination of
exact equations of motion for the mean value of the density
matrix, i.e., the ensemble average over the random fluctuations
for each component $k=1,2,3$. After averaging over the reservoir
variables, we are left with a homogeneous Volterra equation for
the system density operator that has an exponential memory kernel.
The state of the system at time $t$ depends on its past history.
The power spectrum of the environment is given by a Fourier
transform of the exponential correlation function, which is an
unnormalized Lorentzian with a maximum of $2 a^2 \tau$ and a full
width at half maximum equal to $1/(\pi \tau)$.  In the case of
white noise, the delta-function correlation in time leads to a
flat power spectrum for the environment with a strength given by a
diffusion constant. The system is equally coupled to all
frequencies of the external system. In the case of colored noise,
the system prefers certain frequencies. Thus, $a$ is the coupling
strength of the system with the external system while $\tau$
determines which frequencies the system prefers most.  Increasing
both $a$ and $1/ \tau$ corresponds to a more noisy environment.
Therefore, the dimensionless product $a \tau$ is the fundamental
quantity that determines the amount of fluctuation.

Having derived a master equation of the form of Eq.
(\ref{eq:kernelequation}) with an exponential kernel function, we
now determine the damping basis and solve the master equation. We
assume that the fluctuation rates $\tau_i$ are equal so that they
obey the same Poisson statistics.  This leads to a single kernel
operator acting on the Lindblad superoperator, rather than a
linear superposition of such operations: $\dot{\rho}= K_1
\mathcal{L}_1 \rho +K_2 \mathcal{L}_2 \rho+K_3 \mathcal{L}_3
\rho$. The damping basis for Eq. (\ref{eq:rtsrhodot}) is found to
be the following set of eigenvalues and eigenoperators:
$\{\lambda_0, \lambda_1, \lambda_2, \lambda_3
\}=\{0,-4(a_2^2+a_3^2),-4(a_1^2+a_3^2),-4(a_1^2+a_2^2) \}$ and
$\{R_0,R_1,R_2,R_3\}= \{L_0,L_1,L_2,L_3\}=
\{I,\sigma_1,\sigma_2,\sigma_3\}$ which are self dual.  Using the
damping basis, the Laplace transform of Eq. (\ref{eq:rtsrhodot})
becomes
\begin{equation}
    \tilde{\rho}(s)=\sum_i \frac{s+1/\tau}{s(s+1/\tau)-\lambda_i}
    \textrm{Tr}\{ \sigma_i \rho(0) \}  \sigma_i.
\end{equation}
This can be inverted to give the solution
\begin{equation}   \label{eq:solution}
    \rho(t)=\sum_i \textrm{Tr} \lbrace \sigma_i \rho(0) \rbrace
    \Lambda_i(t) \sigma_i.
\end{equation}
In terms of the dimensionless time $\nu=t/ 2\tau$, the functions
$\Lambda_i(\nu)= e^{-\nu} \left[ \cos(\mu_i \nu)+\frac{\sin(\mu_i
\nu)}{\mu_i} \right]$ are damped harmonic oscillators having
frequencies $\mu_i=\sqrt{ (4 \kappa_i \tau_i)^2-1}$ with
$\kappa_i^2= a_j^2+a_k^2$ for $i \neq j \neq k.$ This differs from
the Markovian case, where the functions are purely exponential
functions in time with parameters defining the characteristic
lifetimes.  A power series expansion gives $\Lambda(\nu)=
1-\frac{1}{2}(\mu^2+1) \nu^2+O(\nu)^3$, which shows that the
linear term in $\nu$ is missing. Thus, the standard white noise
diffusion term vanishes.  This is a general property of the memory
kernel and a fundamental difference between white noise and
colored noise.

The function $\Lambda(\nu)$ has two regimes -- pure damping and
damped oscillations.  The fluctuation parameter, given by the
product $\kappa \tau$, determines the behavior of the solution.
When $ 0 \leq \kappa \tau < 1/4$ the solution is described by
damping. The frequency $\mu$ is imaginary with magnitude less than
unity. When $\kappa \tau = 1/4$ the function $\Lambda(\nu)=
e^{-\nu} (1-\nu)$ is unity at the initial time and approaches zero
as time approaches infinity.   In addition to pure damping, damped
harmonic oscillations in the interval $[-1,+1]$ exist in the
regime $\kappa \tau > 1/4$.

The functions $\Lambda_i(\nu)$ determine the evolution of each
component of the Bloch vector. The bound $|\Lambda(\nu)| \leq 1 $
ensures that the density operator evolves only to states on or
inside the Bloch sphere so that $\Phi_t$ always maps positive
operators to positive operators.  For Markovian master equations,
the dissipation results in a contraction of each component, which
is a consequence of the semigroup property.  This property is
absent for colored noise, so that the dissipation results in
contractions with oscillations.  The information exchange between
the system and the environment leads to an exchange of entropy
between the two. The entropy of the average system state can
oscillate in time with an overall decay.

The solution to the master equation (\ref{eq:solution}) defines a
linear map $\Phi_t: \rho \rightarrow \rho_t$ on $\mathcal{M}_2$.
This map has a Kraus decomposition $\Phi_t(\rho) = \sum_k
A_{k}^{\dagger} \rho A_k $ with Kraus operators given by
$A_1=\sqrt{\xi_1(\nu)}\hspace{.1in} \sigma_1,$
$A_2=\sqrt{\xi_2(\nu)} \hspace{.1in}\sigma_2,$
$A_3=\sqrt{\xi_3(\nu)}\hspace{.1in} \sigma_3$, and
$A_4=\sqrt{\xi_4(\nu)}\hspace{.1in} I,$ provided the following
linear combinations are nonnegative:
\begin{eqnarray}  \label{eq:econditions}
    4\xi_1(\nu)&=&1+\Lambda_1-\Lambda_2-\Lambda_3 \geq 0,  \nonumber \\
    4\xi_2(\nu)&=&1-\Lambda_1+\Lambda_2-\Lambda_3 \geq 0,  \\
    4\xi_3(\nu)&=&1-\Lambda_1-\Lambda_2+\Lambda_3 \geq 0,  \nonumber  \\
    4\xi_4(\nu)&=&1+\Lambda_1+\Lambda_2+\Lambda_3 \geq 0.  \nonumber
\end{eqnarray}
We now show which properties (2) hold. Clearly, this mapping is
trace-preserving and property 2$(i)$ is satisfied. Property
2$(iii)$ is satisfied because $\textrm{lim}_{t \downarrow 0}
\Lambda_i(t)= 1$. The system evolves continuously in time and the
evolution is described by the identity map at the initial time.
The semigroup property 2$(iv)$ is lost as $\Lambda_i(t)
\Lambda_i(s) \neq \Lambda_i(t+s)$.

The map is completely positive if $\Phi_t \otimes I_n \geq 0
\hspace{.05in} \forall \hspace{.05in} n \in \textrm{Z}_+$. It is
sufficient to show that the composite operation on a maximally
entangled state is positive \cite{choi1972}. For a linear map from
$\mathcal{M}_2$ to $\mathcal{M}_2$, we need only show that the
composite map $\Phi_t \otimes I \left( | \beta_{00} \rangle
\langle \beta_{00} | \right)$ on $\mathcal{M}_4$ is positive
semidefinite, where $| \beta_{00} \rangle =\frac{1}{\sqrt{2}}(|00
\rangle + |11 \rangle)$ is the maximally entangled Bell state.
Hence the map is completely positive if and only if the
eigenvalues of the composite map are nonnegative.  The eigenvalues
$\{\xi_j \}$ are exactly those given by Eqs.
(\ref{eq:econditions}). Therefore, property 2$(ii)$ is satisfied
for all $a_i \tau$ and every $\nu$ if and only if
\begin{equation}  \label{eq:infimum}
    \inf_{\left(a_i \tau \right)} \xi_j(\nu) \geq 0, \hspace{.2in}
    j=1,2,3,4.
\end{equation}
This follows from the fact that the existence of a Kraus
decomposition and the composite operation on the maximally
entangled state are both necessary and sufficient to show complete
positivity.

Condition (\ref{eq:infimum}) is not satisfied for all values of
the parameters. A case where it is satisfied occurs when two or
more of the $a_i$ are zero.  For example, if two of the $a_i$ are
zero and one is nonzero then $\Phi_t$ defines a depolarizing
channel with colored noise. Suppose $a_3=a, a_1=a_2=0$. In this
case, $\Lambda_1(\nu)=\Lambda_2(\nu)=\Lambda(\nu)$ and
$\Lambda_3(\nu)=1.$ This map has Kraus operators given by
$A_1=\sqrt{[1+\Lambda(\nu)]/2}\hspace{.1in} I$ and
$A_2=\sqrt{[1-\Lambda(\nu)]/2}\hspace{.1in} \sigma_3$.  This map
is a completely positive, trace-preserving map for all values of
the fluctuation parameter $a \tau$ because $|\Lambda(\nu)| \leq 1
$ for all times $\nu$.  The Hamiltonian $H=\hbar \Gamma_3
\sigma_3$ implies that the $z$ component of the spin-1/2 particle
is a constant of the motion so that the two states $| \pm
\rangle_z$ are fixed points and do not evolve.
    As a physical model, a constant magnitude magnetic field is
applied in the $z$ direction which inverts randomly in time.  If
the field is unobserved, the average trajectory for the density
operator of the spin-1/2 system is dissipative.  As time
approaches infinity, there is maximum uncertainty in the $x$ and
$y$ components of the spin system.  A steady state
$\Phi_t(\rho)_{ss}=\frac{1}{2}(\rho+\sigma_3 \rho \sigma_3)$ is
reached; the entire Bloch sphere evolves to a line connecting the
north and south poles.

An example where condition (\ref{eq:infimum}) fails is given by
$a_1=a_2=a$ and $a_3=0$. Then each component of the Bloch vector
evolves with frequencies given by $\mu_1=\mu_2=\sqrt{(4a
\tau)^2-1}$ and $\mu_3=\sqrt{32(a \tau)^2-1}$.  There are no fixed
points except the maximally mixed state.  In this case, it is
found numerically that $a \tau < 0.8$ leads to a completely
positive map.  This means that there is a limit on the environment
power spectrum. The bandwidth may be large provided the coupling
is made small and vice versa.

We find that if two or more of the $a_i$ are nonzero then there
are regimes for the fluctuation parameters where complete
positivity is lost.  Assume the frequencies $\mu_i$ are equal.
Then only the last eigenvalue in (\ref{eq:econditions}) need be
considered.  Setting the time equal to $\nu=\pi/\mu$ we find that
if the frequencies are less than or equal to $\pi/ln(3)$ the map
defined by Eq. (\ref{eq:solution}) is completely positive for all
time. Complete positivity is lost as the frequencies become large.
The case of three equal frequencies sets an upper bound.  We have
the following sufficient condition for (\ref{eq:infimum}) to be
satisfied:  \textit{If} $\mu^\star=\textrm{max}\{\mu_1,\mu_2,\mu_3
\}\leq \pi/ln(3)$ \textit{then the map is completely positive for
all time.}

We recover the Markovian master equation by letting $\tau
\rightarrow 0$ and $a \rightarrow \infty$ in such a way that $2
a^2 \tau$ becomes a constant.  The random telegraph signal reduces
to a Gaussian white noise in this singular limit and Eq.
(\ref{eq:rtsrhodot}) becomes
\begin{equation}
    \dot{\rho}(t) = -  \int_0^t \delta(t-t') \sum_k  2 a^2_k \tau
    [\sigma_k,[\sigma_k, \rho(t') ]] \textrm{d}t'.
\end{equation}
This master equation is local in time, leading to
$\Lambda_i(t)=\exp(-\gamma_i t)$ in Eq. (\ref{eq:solution}) with
inverse lifetimes $\gamma_i = 4 \kappa_i^2 \tau$.  We point out
that, even in the case of white noise, there are examples of maps
that are positive but not completely positive
\cite{gorini1976,kimura2002,daffer2003}.  For Markovian dynamics,
the relations (\ref{eq:econditions}) are satisfied if and only if
$\gamma_i \leq \gamma_j + \gamma_k$ for all permutations of the
indices.

By construction of our example, complete positivity holds in the
white noise limit. We conclude that the loss of complete
positivity, when the memory kernel is present, is a feature of the
colored noise.  The validity of Eq. (\ref{eq:kernelequation}) has
been questioned \cite{barnett2001}. This Letter shows conditions
under which a memory kernel is valid and shows that such
conditions can be achieved. We have presented an example of a
completely positive, trace-preserving map that results from a
system-environment coupling that is essentially non-Markovian.
From this, it was shown that a memory kernel can be physically
valid. The results reveal that interesting features arise from the
colored noise of the environment, which are not present in the
limit of white noise. White noise is an idealization of real
noises and under certain conditions cannot be used.  Thus, more
work needs to be done in the study of more general noises that
include memory effects.

This work was partially supported by a KBN grant No. 2PO3B 02123
and the European Commission through the Research Training Network
QUEST.


\end{document}